\documentclass[11pt,a4]{article}
\usepackage{amsmath}
\usepackage{graphicx}
\begin{document}

\title{Analytical study of the structure of chaos near unstable
points}
\date{}
\author{G. Contopoulos \thanks{gcontop \@ academyofathens.gr}, C. Efthymiopoulos  \& M. Katsanikas  \\ Research Center for Astronomy, Academy of Athens \\ Soranou Efesiou 4, GR-115 27 Athens, Greece}

\maketitle

\vskip 1cm

\section{Abstract}
In a 2D conservative Hamiltonian system there is a formal integral
$\Phi$ besides the energy H. This is not convergent near a stable
periodic orbit, but it is convergent near an unstable periodic
orbit. We explain this difference and we find the convergence radius
along the asymptotic curves. In simple mappings this radius is
infinite. This allows the theoretical calculation of the asymptotic
curves and their intersections at homoclinic points. However in more
complex mappings and in Hamiltonian systems the radius of
convergence is in general finite and does not allow the theoretical
calculation of any homoclinic point. Then we develop a method
similar to analytic continuation, applicable in systems expressed in
action-angle variables, that allows the calculation of the
asymptotic curves to an arbitrary length. In this way we can study
analytically the chaotic regions near the unstable periodic orbit
and near its homoclinic points.

\section{Introduction}

It is well known that the formal integrals near stable periodic
orbits are not convergent. Nevertheless, their finite truncations
 represent with remarkable accuracy the quasi-periodic motions
near the stable periodic orbits(for a review see Contopoulos
$2002$)\cite{Cont2002}. On the other hand, it is less well known
that the formal integrals are convergent near \textbf{unstable}
periodic orbits.

In order to explain this difference,  let us consider the case of
two coupled oscillators with Hamiltonian

\begin{equation}
H=\frac{1}{2}(p_1^2+ \omega ^2_1 x^2_1)+
\frac{1}{2}(p^2_2+\omega_2^2 x^2_2)+ H_3 + H_4 +...
\end{equation}
where the frequencies $\omega_1,\omega_2$ are real the ratio
$\frac{\omega_1}{\omega_2}$ is irrational. A particular formal
integral near the origin $p_2=x_2=p_1=x_1=0$ is of the form

\begin{equation}
\Phi=\Phi_2+\Phi_3+\Phi_4+
\end{equation}
where $\Phi_2=\frac{1}{2}(\dot{x}^2_1+\omega^2_1x^2_1)$ and its
higher order terms are of the form

\begin{equation}
\Phi_s= \Sigma \frac{x_1^{s_1} p_1^{s_2} x_2^{s_3} p_2^{s_4}}
 {m \omega_1+n\omega_2}
\end{equation}
where $s_1+s_2+s_3+s_4=s$ and $|m|+|n|=s$. The divisors $m
\omega_1+n \omega_2$  can become arbitrarily small for particular \emph{m} and \emph{n }, and this leads to the nonconvergence of the series$(2)$.

\ \ On the other hand if one frequency, say $\omega_2$, is imaginary
$(\omega_2= - i\nu)$ the divisors never approach zero and the series $(2)$ is convergent in a certain domain around the origin. Near the origin, in this case, we have chaos. But we can use these convergent series to study the chaotic motions analytically.

\ \ The convergence of the integrals near unstable periodic orbits
was first demonstrated by Moser ($1956$\cite{Mo1956},$1958$
\cite{Mo1958} ) and the proof was completed by Giorgilli
($2001$)\cite{Gio2001}. Da Silva Ritter et al.
($1987)$\cite{DaSi1987} demonstrated that in simple 2D mappings the
domain of convergence extends to infinity along the asymptotic
curves emanating from the unstable points.

\ \ These asymptotic curves intersect at an infinity of homoclinic points (Fig.1). The curves form elongated oscillations and cover the neighborhood of the unstable point \emph{O }where chaos is dominant. Thus the formal series can be used to describe the chaotic domain close to the unstable point \emph{O} .

\ \ On the other hand in the case of Hamiltonian systems of two
degrees of freedom some numerical results up to now Vieira et
al.($1996$)\cite{Vi1996}), and Bongini et al. ($2001$)\cite{Bong2001}) indicate that the domain of convergence extends at most up to  the central homoclinic point $H_o$(opposite to \emph{O} in Fig.1).

\ \ In the present paper we briefly summarize our results of
Efthymiopoulos et al.($2013$)\cite{Efthym2013}regarding: i)the
problem of the convergence both in the case of mappings and in the
case of Hamiltonian systems,  and ii) a new method for extending the
calculation of the invariant curves arbitrarily beyond the central
homoclinic point in the Hamiltonian case. For details see
Efthymiopoulos et al.($2013$)\cite{Efthym2013}

\begin{figure}
\begin{center}
\includegraphics*[width=.6 \textwidth]{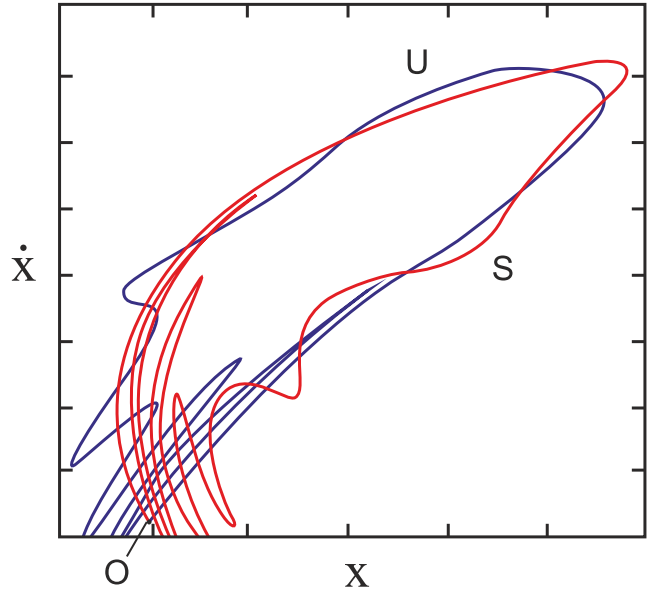}
\caption{ The unstable (U) and stable (S) asymptotic curves from the unstable invariant point O intersect at an infinity of homoclinic points (schematically).}
\end{center}
\end{figure}


\section{Mappings}
We consider a real analytic 2D mapping of the form
\begin{eqnarray}
x_1^{'}&=&\lambda_1x_1+F_2(x_1, x_2)+...\\
x_2^{'}&=&\lambda_2 x_2+G_2(x_1 ,x_2)+...\nonumber
\end{eqnarray}
where $\lambda_1=e^\nu,\lambda_2=e^{-\nu} (\nu>0).$

\ \ Then we find a canonical transformation to new variables
$(\xi,n)$ such that the mapping takes the normal form
\begin{eqnarray}
\xi^{'}&=&W_1(\xi,\eta)=\Lambda(c)\xi\\
\eta^{'}&=&W_2(\xi,\eta)=\frac{1}{\Lambda(c)}\eta\nonumber
\end{eqnarray}
where
\begin{equation}
\Lambda(c)=\lambda_1+w_{1} c+w_{2} c^2+...
\end{equation}
with
\begin{equation}
c=\xi \eta = \xi{'} \eta{'}
\end{equation}
The asymptotic curves from the origin are $\xi=0$ and $\eta=0.$ As
particular examples we consider
 (a) the standard map
\begin{eqnarray}
x_1^{'}&=& x_1 +K \sin(x_1 + x_2)\\
x_2^{'}&=&x_1 + x_2 \nonumber
\end{eqnarray}
 which is of the form $(4)$ after a linear diagonaling transformation, and (b) the H\'{e}non map
\begin{eqnarray}
x_1^{'}&=&e^a[x_1-\frac{1}{4}(x_1+x_2)^2]\\
x_2^{'}&=&e^{-a}[x_2+\frac{1}{4}(x_1+x_2)^2]\nonumber
\end{eqnarray}
\ \ In both cases we truncate the series $\Lambda(c)$, giving $\xi$ and $\eta$, at a particular order N and then we transform back to the original variables $x_1,x_2$ in order to find the invariant curves $\xi=0$ and $\eta=0$ in the variables $x_1,x_2.$


\begin{figure}
\begin{center}
\includegraphics*[width=.9 \textwidth]{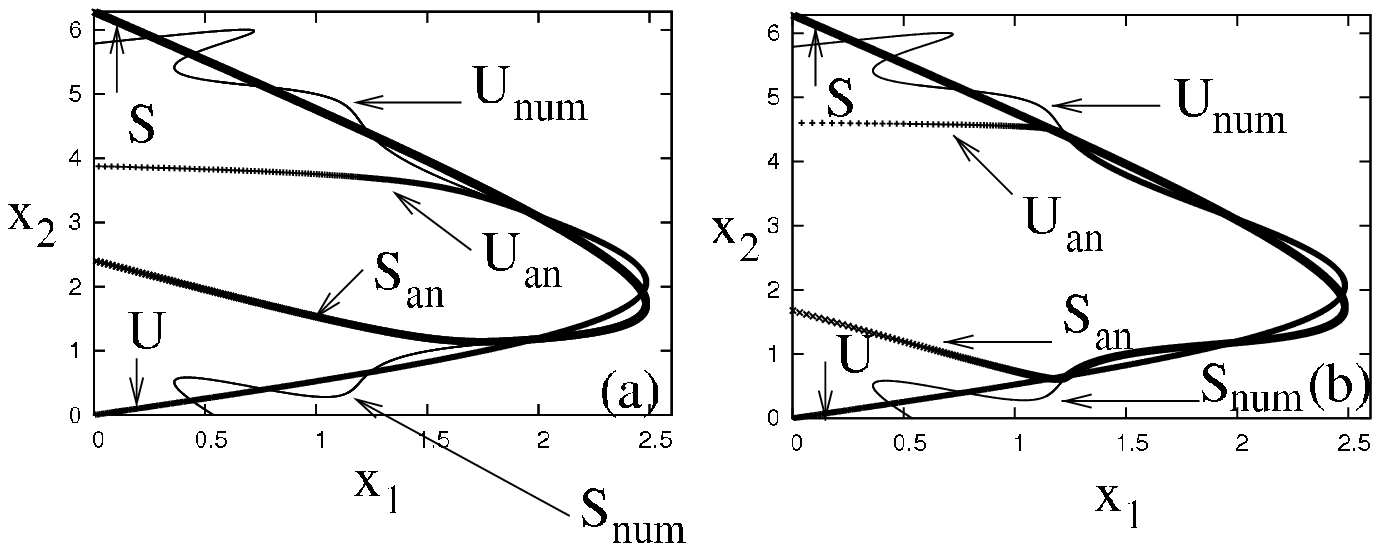}
\caption{The accurate (numerical) asymptotic curves $(U_{num}$ from the invariant point $x_1=x_2=0$, and $S_{num}$ from $x_1=0,x_2=2\pi$) and the theoretical (analytical) curves $U_{anal}$ and $S_{anal}$ of the standard map for $K=1.3$ truncated at orders (a)$N=20$ and (b)$N=60.$}
\end{center}
\end{figure}


\ \ In the case of the standard map we start the unstable invariant curve at the point $(x_1=x_2=0$) and the stable invariant curve at the point $(x_1=0,x_2=2\pi)$. In Fig.2 we draw the numerical results by thin lines and the analytical results by thick lines. If we truncate the series at order $N=20$ we find the thick curves of Fig.2a. We see that the analytic curves agree with the numerical results up to one homoclinic point $(H_1)$ beyond (and before $(H_{-1}$)) the central homoclinic point $H_0$. But beyond that point the analytical curves $U_{anal}$ and $S_{anal}$ deviate from the numerical curves $U_{num}$ and $S_{num}.$

\ \ When the truncation takes place at order $N=80$ (Fig.$2b$) the analytic curves agree with the numerical curves up to the second homoclinic point $(H_2)$ beyond $H_o$ or before it $(H_{-2})$.

\ \ In the case of the H\'{e}non map both unstable and stable asymptotic curves start at $(x_1=x_2=0)$. In this case if we truncate the series at order $N=20$ we find agreement between the analytical and numerical curves ($U_{anal},S_{anal}$ and $U_{num},S_{num}$) up to the second homoclinic point beyond $H_o$ (Fig.$3a$). If we truncate the series at the order $N=60$ we find agreement even beyond the fourth homoclinic point $(H_4)$ beyond $H_o$ (Fig.$3b$). In this latter case we find that the analytic asymptotic curves come quite close to the original unstable point \emph{O}, thus they enter in the chaotic region near \emph{O}.

\begin{figure}
\begin{center}
\includegraphics*[width=.9 \textwidth]{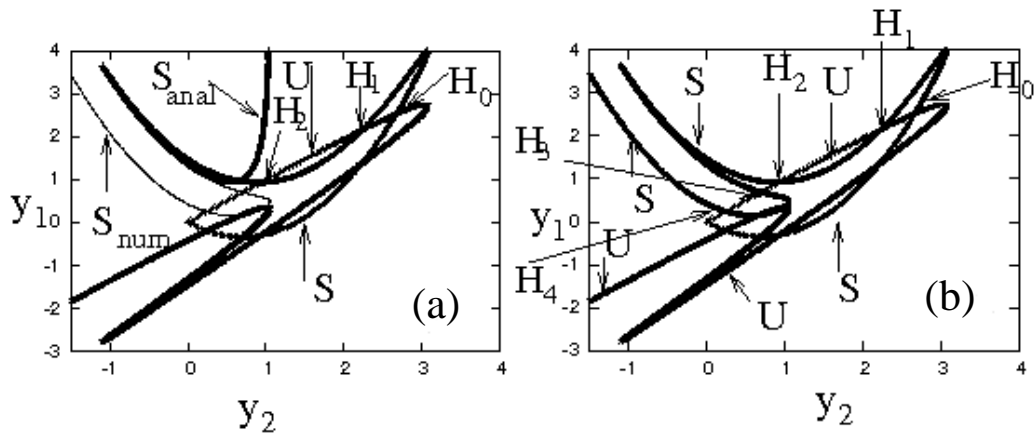}
\caption{The same curves as in Fig.2 emanating from the invariant point \emph{O} in the case of the H\'{e}non map, for $\alpha=1.43$. truncated at orders (a) $N=20$ and (b)$N=60$.}
\end{center}
\end{figure}


\ \ It can be shown that in both the standard map and the H\'{e}non
map the radii of convergence of the series along the asymptotic
curves are infinite \cite{DaSi1987}. This can be demonstrated
numerically by using the d' Alembert criterion.i.e. by calculating
the absolute value of the ratio of the successive terms of the
series $\rho_N=|\Phi_N|/|\Phi_{N-1}|$. We find that this
 increases with N. However the approach
to infinity is quite different in the standard map and in the
H\'{e}non map. Namely $\rho_N$ increases linearly in $\log N$ in the
standard map, while it increases like $N^2$ in the H\'{e}non map.
Thus the increase is much slower in the case of the standard map
than in the case of the H\'{e}non map. This explains why in Fig.2
the series at order $N=80$ represent the true (numerical) results
only up to the second homoclinic point beyond $H_o$ in the first
case, while the series at order $N=60$ reach the fourth homoclinic
point in the second case.

\ \ However the convergence radius is infinite only if the mapping is analytic over the whole domain covered by the invariant manifolds. Otherwise the convergence radius is finite. As an example we consider the mapping

\begin{eqnarray}
x^{'}_1&=&x_1+\frac{K\sin(x_1+x_2)}{2-\cos(x_1+x_2)}\\
x^{'}_2&=& x_1+x_2\nonumber
\end{eqnarray}
which has the same unstable point (the origin)and the same
eigenvalues and eigenvectors as the standard map . However if we
develop the formula for $x^{'}_1$ as a Taylor series we find that it
has a singular points for $\cos(x_1+x_2)=2$, i.e. for imaginary
values of $x_1+x_2=\pm 1.31696 i$. Therefore it converges only if
$|x_1+x_2|<1.31696$. For larger $|x_1+x_2|$ the hyperbolic normal
form fails to represent the numerical invariant curves.In fact it
does not even reach the central homoclinic point (Fig.4).

\begin{figure}
\begin{center}
\includegraphics*[width=.8 \textwidth]{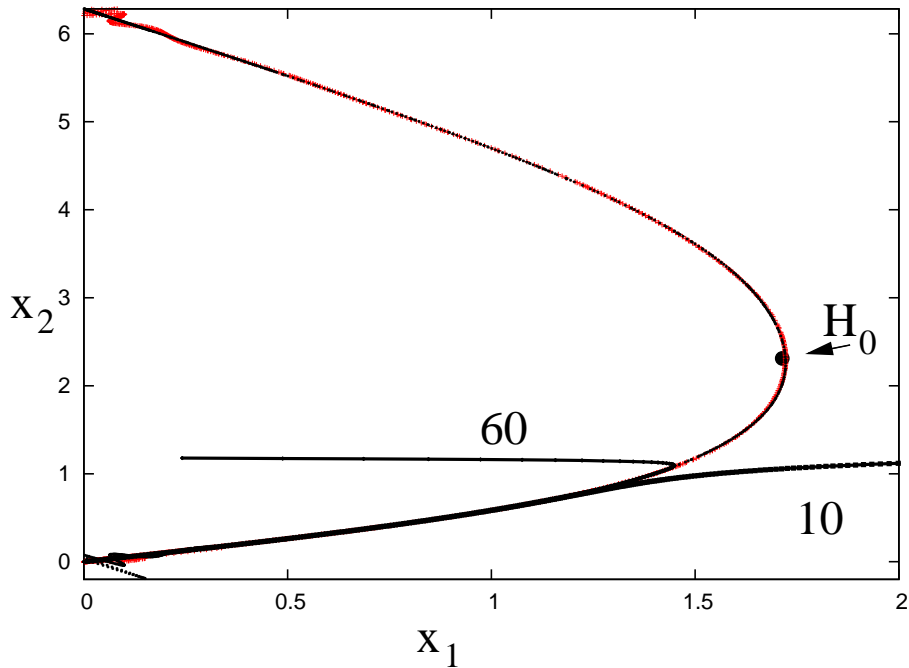}
\caption{The numerical asymptotic curves U (thin red, from the invariant point $x_1=x_2=0)$ and S (thin black, from $x_1=0,x_2=2\pi$) in the case of the mapping(10) with $K=1.3$ intersect at several homoclinic points, like $H_o$. These curves make conspicuous oscillations only near the invariant points. On the other hand the analytical curves from $x_1=x_2=0$, truncated at orders $N=20$ and $N=60$ (thick black curves) are close to the numerical curves up to a certain distance, approaching the central homoclinic point $H_o$ closer as the order increases, but not reaching it.}
\end{center}
\end{figure}



\section{Hamiltonians}

We consider the Hamiltonian (Efthymiopoulos $2012$
\cite{Efthym2012b})

\begin{equation}
H=\frac{p^2}{2}-\omega^2_o[1+c(1+p)\cos\omega t]\cos\psi
\end{equation}
which is  $1 \tfrac{1}{2}$ degrees of freedom. The Hamiltonian (11)
represents a perturbed pendulum. In fact, if $\varepsilon=0$ we have
the pendulum Hamiltonian $H_o=\frac{p^2}{2}-\omega_ o^2 \cos\psi$.
The Hamiltonian (11) can be written as a 2- dimensional Hamiltonian

\begin{equation}
H=\frac{p^2}{2}+\omega I -\omega^2_o[2+\varepsilon(1+p)\cos
\varphi]\cos \psi
\end{equation}
if we introduce a dummy action \emph{I}, conjugate to the angle
$\varphi=\omega t.$

\ \ We calculate orbits in this Hamiltonian (12) and we find a Poincar\'{e} section (Fig.$5a$) by plotting the points $(\psi, p)$ at successive times $ t=nT=n \frac{2\pi}{\omega} (n=1,2...).$ The points are given in the interval $-\pi\leq \psi \leq \pi$, and they are repeated modulo $2\pi$ for longer and smaller $\psi.$

\ \ The main part of Fig.$5a$ contains chaotic orbits, extending all over the interval $-\pi\leq \psi \leq \pi$. However there are two large islands of stability around two stable orbits on the $\psi=0$ axis. Furthermore there are invariant curves from $-\pi$ to $+\pi$ above and below the chaotic zone, and small secondary islands.


\begin{figure}
\begin{center}
\includegraphics*[width=\textwidth]{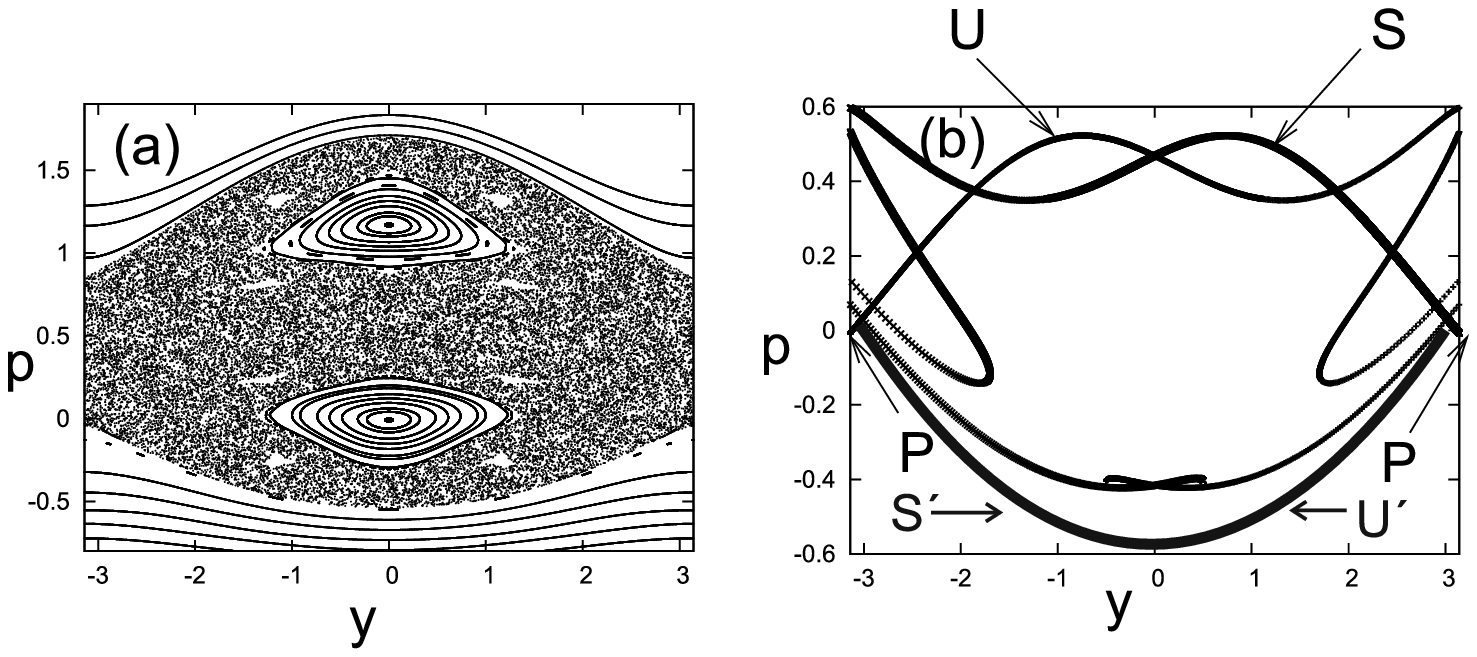}
\caption{(a) The distribution of the orbits on the Poincar\'{e} surface of section for the Hamiltonian (12) with
$\varepsilon=\omega=2,\omega_o=0.2\sqrt{2}$(successive points in the plane $(\psi,\emph{p})$ at times
$t=nT=n\frac{2p}{\omega}(n=1,2...).$ (b) The asymptotic curves from $P(y=-\pi)$ (U unstable and $S^{'}$ stable)and from $P^{'}(y=\pi)$ (S stable and $U{'}$ unstable).}
\end{center}
\end{figure}


\ \  The chaotic region is generated by the unstable periodic orbit $\emph{P}$ which has $\psi=-\pi$ (equivalent to $\psi=\pi$) and $p=p_o$ close to zero. In fact in Fig. $5a$ we have $p_o = -0.0073... .$ The initial unstable (U) and stable (S) asymptotic curves of the point $\emph{P}$ are shown in Fig.$5b$, and if we extend them they cover all the chaotic domain.

\ \ If we expand the $\cos \psi$ term we have

\begin{equation}
\cos \psi=\cos (\pi +u)=-1 +\frac {u^2}{2} +...
\end{equation}

and we write also

\begin{equation}
 \cos \varphi= \frac {1}{2} (e^{i\varphi} + e^{-i \varphi})
\end{equation}

The Hamiltonian (12) is a perturbation of the hyperbolic Hamiltonian

\begin{equation}
H_h = \frac {p^2}{2} - \nu^2 \frac{u^2}{2}
\end{equation}

where $ \nu = real = i \omega.$

If we introduce canonical variables $(\xi, \eta)$ such that
\begin{equation}
u = \frac{1}{\sqrt{2\nu}}(\xi-\eta),  p=\frac{\sqrt{\nu}}{\sqrt{2}}
(\xi +\eta)
\end{equation}

we write the Hamiltonian in the form

\begin{equation}
H=\omega I+\nu\xi\eta+H_1 (\varphi,I,\xi,\eta)
\end{equation}

Then we use the method of Lie series (Giorgilli $2001$\cite{Gio2001}, Efthymiopoulos $2012$\cite{Efthym2012b}) to bring the Hamiltonian in normal form using new canonical variables $\xi{'},\eta{'}.$ In this way the Hamiltonian  H becomes a function of only the product $c=\xi^{'}\eta^{'}$ and \emph{I}. The quantity $c=\xi^{'}\eta^{'}$ is now an integral of motion in the final variables $\xi^{'},\eta^{'}.$

\ \ However, in practice, the normalization of the Hamiltonian takes place at a finite order  $r_{max}$, e.g. (a) $r_{max}=40$ and (b) $r_{max}=80.$ In fact we compute the analytic series for $\xi^{'}=0$, or $\eta^{'}=0$ up to order $r_{max}$, and returning to the original variables $(\psi,p)$ we find the asymptotic curves U from $P(\psi=-\pi)$ and S from $P(\psi=\pi)$ and compare them to the numerical asymptotic curves.


\begin{figure}
\begin{center}
\includegraphics*[width=.4 \textwidth]{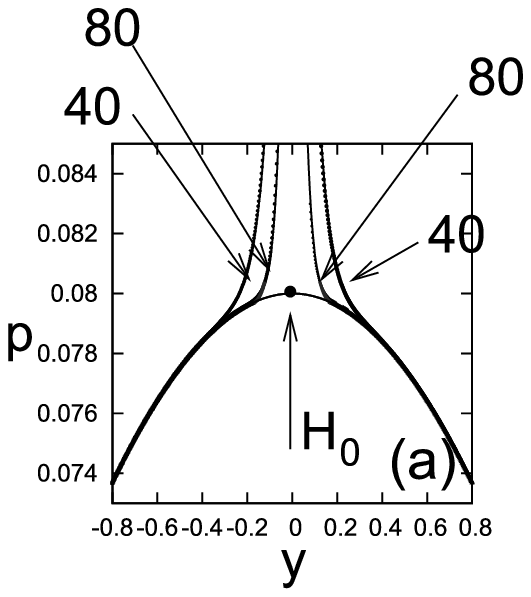}
\caption{The accurate asymptotic curves U and S (thin lines) in the pendulum case  $(\psi=0)$ form a continuous curve, passing through the middle point $H_o$ (the limit of the homoclinic point when $(\varepsilon\rightarrow 0).$ The analytical curves truncated at orders $r_{max}=40$ and $r_{max}=80$ (thick lines) approach $H_o$ as the order increases, but they do not reach it.}
\end{center}
\end{figure}



\begin{figure}
\begin{center}
\includegraphics*[width=.6 \textwidth]{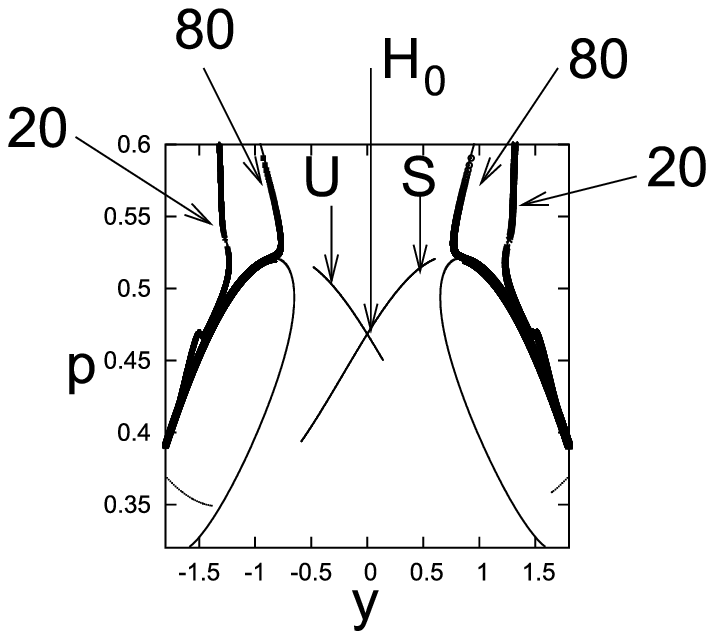}
\caption{The same as in Fig.6 for the case $\varepsilon=1$.}
\end{center}
\end{figure}


\ \ A first application is in the case $\varepsilon=0$ (the case of the pendulum). The exact (numerical) asymptotic curves U an S from $P(\psi=-\pi)$ and $P=(\psi=\pi)$ join smoothly at the point $H_o,$ where $\psi=0$ (Fig.$6$). However the analytic expansions at order $r_{max}=20$ and $r_{max}=80$ do not reach the point $H_o$, although they approach it better for larger $r_{max}$. For much larger $r_{max}$ the theoretical curves come even closer to $H_o$, but without reaching it. Thus the convergence of the analytic series is only up to the point $H_o$. It is remarkable that even in this simple case (the pendulum) the analytic series apply only up to the middle point $H_o.$

\ \ Similar results are found for $\varepsilon=1$ (Fig.$7$). The analytic curves apply only to parts of the real (numerical) asymptotic curves and do not even reach the central homoclinic point $H_o$.

\ \ Similar indications that the convergence of the analytic (normal form) series reach only the central homoclinic point $H_o$ have been found by Vieria et al.($1996$)\cite{Vi1996} and by Bongini et al. ($2001$)\cite{Bong2001}.


\section{Extended method}

\ \ In order to find series applicable beyond the central homoclinic point $H_o$ we use action angle variables and not cartesian coordinates as in the previous sections. In particular, in Eq.($12$)we do not expand $\cos \psi$ in Taylor series.

\ \ More generally we use Fourier series in the angles, and we
exploit the fact that the Fourier coefficients decay exponentially
(Giorgilli $2001$.\cite{Gio2001}).

\ \ If we consider $\psi$ as a complex variable the domain of
analyticity covers the whole real interval $(0\leq Re (\psi)\leq
2\pi)$ but it has a limited imaginary extent $( (Im (\psi)\leq
\sigma )$ (where, $\sigma > 0$). Thus we cannot have a large radius
of convergence for $\psi$. However, if we split the  interval $(0,2
\pi)$ of the angle $\varphi$ in parts where the convergence is
guaranteed, by separating the  corresponding times $(O,
T=\frac{2\pi}{\omega})$  into intervals $\Delta t_1, \Delta t_2 ...
\Delta t_n$ such that $(\Delta t_1 +\Delta t_2 +...+\Delta_n)$=
\emph{T}, then, the composition of the series produced by
analytically continuing from one interval to the other are
convergent.

\ \ The asymptotic curves are the axes $\eta^{'}=0$ (unstable) and $\xi^{'}=0$ (stable) in the new variables (Fig.$8$), while the corresponding curves in the original variables $(\xi,\eta)$ are the curves U and S of Fig.$8$. In order to find the mapping of any point $\alpha$ of the axis $\eta^{'}=0$  to the corresponding point (A) in Fig.$8$ we proceed as follows : $(1)$ We calculate the $m-th$ pre-image (at time $t=-m T$) along the axis $\eta^{'}=0$ and reach a point $\alpha^{'}$, at sufficiently small distance from O, so that the mapping to the old variables ( to the point $A^{'}$) is convergent. $(2)$ We calculate the $mth$ image of the point $A^{'}$,(separating every interval \emph{T} into pieces, as indicated above) and find the point A, which is the image of the point $\alpha$.


\begin{figure}
\begin{center}
\includegraphics*[width=.5 \textwidth]{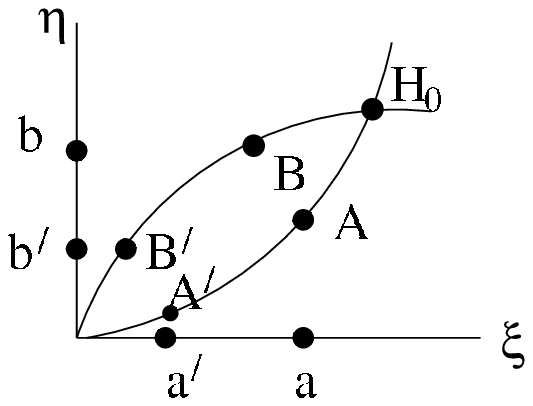}
\caption{The asymptotic curves U and S in the variables $(\xi,\eta)$ intersect at the homoclinic point $H_o$, while these asymptotic curves in the variables  $(\xi^{'},\eta^{'})$ are the axes $\eta^{'}=0$ and $\xi^{'}=0.$ The point $\alpha$ on the axis $\eta^{'}=0$ has a preimage $\alpha^{'}$ close to the invariant point O. The corresponding points in the old variables $(\xi~\eta)$ are A and $A^{'}$ on the curve U. Similarly the images of b and $b^{'}$ are B and $B^{'}$ on the curve S.}
\end{center}
\end{figure}


\ \ We repeat the same process along the stable manifold $\xi^{'}=0$ to find the image B of the point $\beta$. These processes are exact if the original segments $(O\alpha)$ and $(O\beta)$ are not very long.

\ \  Applying this method to the Hamiltonian $(12)$ we found an appropriate separation of T in $4$ pieces $\Delta t_1=\Delta t_2=\Delta t_3=\Delta t_4= T/4.$ Then by taking $m=1$ we found the bold curves of Fig.$9a$. Namely we calculated the unstable asymptotic curve U from the periodic orbit $P(\psi=-\pi)$ and the stable asymptotic curve S from $P(\psi=\pi)$, and found that the theoretical curves U and S agree with the exact (numerical) curves up to beyond the homoclinic points $H_1$ and $H_{-1}$ (Fig.$9a$).


\begin{figure}
\begin{center}
\includegraphics*[width=.8 \textwidth]{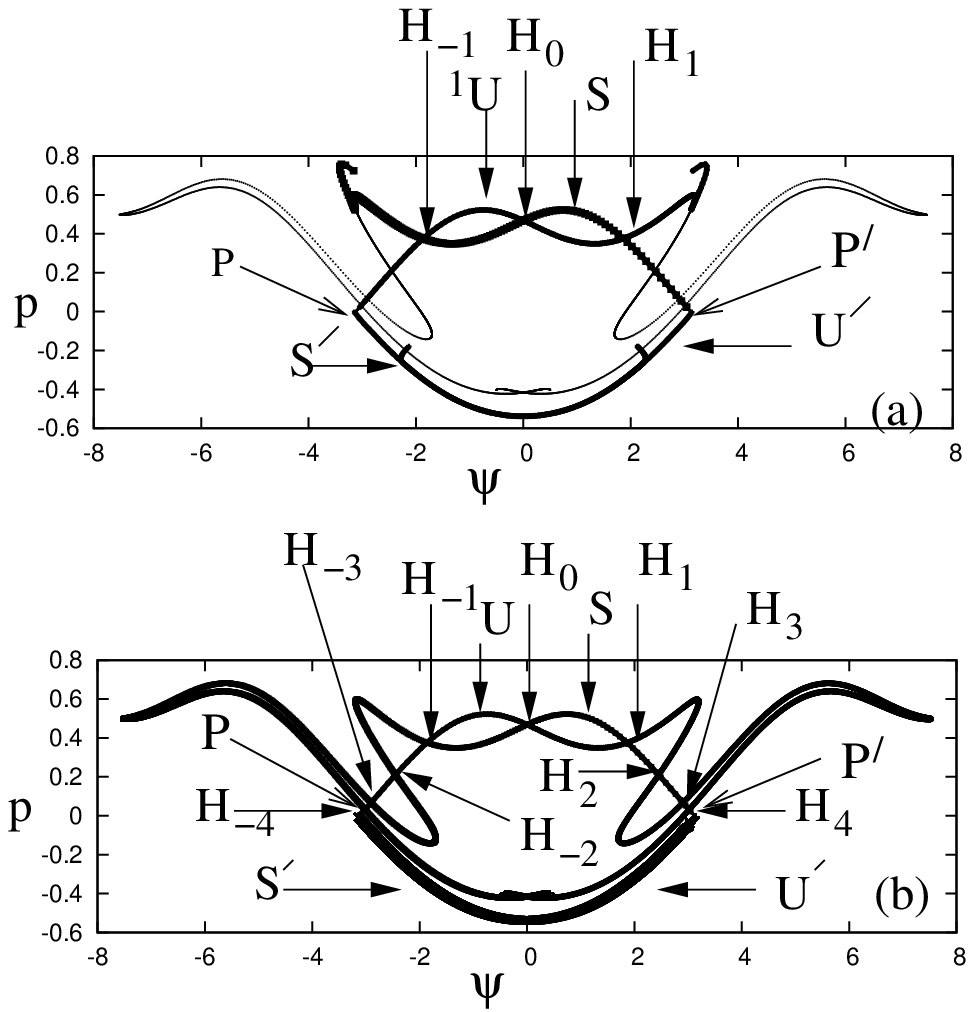}
\caption{The numerical asymptotic curves U from the point $P(\psi=-\pi)$ and S from $P^{'}(\psi=\pi)$ (thin lines) intersect at various homoclinic points $H_o$, etc. The analytical curves calculated by our new method are represented by thick lines by taking (a)$m=1$ and (b)$m=3$. In the first case the analytical curves agree with the numerical ones up to the points $H_1, H_{-1}$. In the second case they agree up to the points $H_4, H_{-4}.$ The asymptotic curves $S^{'}$ from $P(\psi=-\pi)$ and $U^{'}$ from $P(\psi=\pi)$ have similar intersections but the amplitude of their oscillations is very small.}
\end{center}
\end{figure}


\ \ Doing the same calculations with $m=3$, we found agreement
between the analytical and numerical results up to beyond the
homoclinic points $H_4, H_{-4}$ (Fig.$9b$). We can then find the
homoclinic points  to a high accuracy by using a Newton-Raphson
method. In Fig.$9b$ we see that the analytical asymptotic curves
make large oscillations close to the unstable invariant  points.
Thus they enter in the chaotic regions around these points. By
extending the asymptotic curves (U and S) even further we find that
their oscillations reach also the chaotic regions near the
homoclinic points.

\ \ In a similar way we find the intersections of the asymptotic curves $U^{'}$ from $\emph{P}(\psi=\pi)$ and $S^{'}$ from $\emph{P}(\psi=-\pi)$ (Figs.$9a,b$). However in this case the oscillations of the curves $U^{'}$ and $S^{'}$ are very small.

\ \ Similar results are found in more general Hamiltonians if they are expressed in action-angle variables.

\ \ The Hamiltonian cases that we considered in sections $3$ and $4$ are very similar to the mappings of section $2$. In fact in the Hamiltonian case we calculate the mappings along the asymptotic curves starting with points on the axes of the new variables $(\xi^{'}, \eta^{'})$ and we find the corresponding points on the asymptotic  curves U,S ( or $U^{'},S^{'}$) in the old variables.

\ \ However there is an important difference between the simple
mappings, like the standard map and the H\'{e}non map, and the
mappings generated by the Hamiltonians. In the simple mappings of
section 2 the series transformations have an infinite radius of
convergence, while in the Hamiltonian case the radius of convergence
of the series is finite. This is why, by using Taylor series, as in
section 3, we cannot even find the central homoclinic point $H_o.$
On the other hand, by the method exposed above we obtain a
representation of the

\begin{figure}[ht]
\begin{center}
\includegraphics*[width=.8 \textwidth]{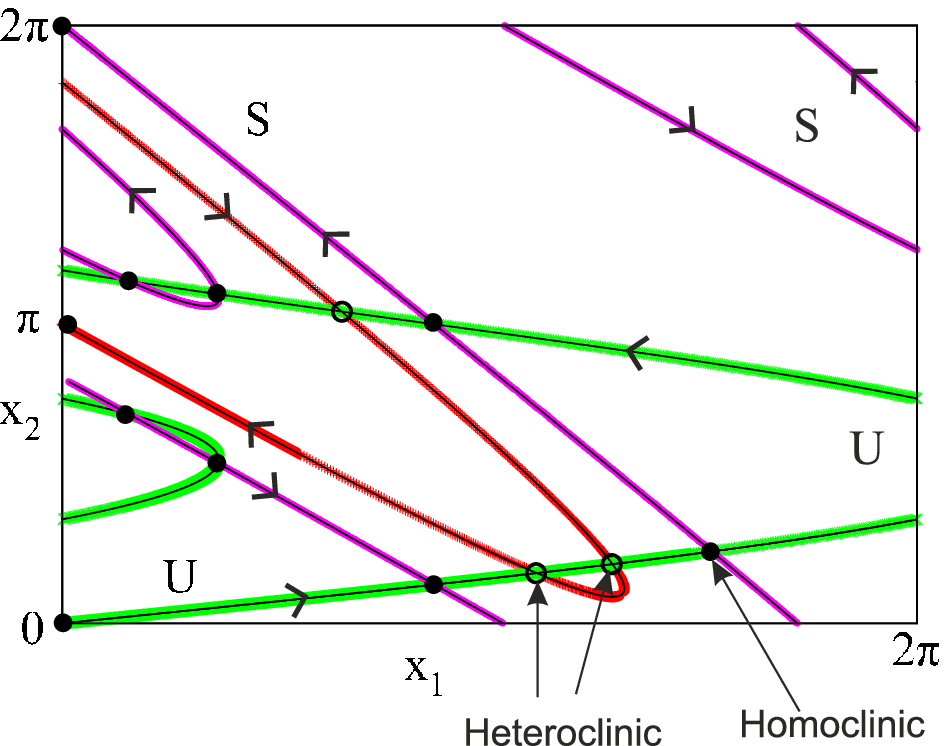}
\caption{Asymptotic manifolds in the standard map for $K=6$. The unstable manifold of the orbit $(x_1=x_2=0)$ (green) intersects the corresponding stable manifold (magenta) at several homoclinic points $(\bullet)$ and the stable manifold from the orbit $(x_1=0,x_2=\pi)$(red) at 3 heteroclinic points $(o)$.}
\end{center}
\end{figure}
Poincar\'{e} mapping of $(\xi(0),\eta(0))$ to
$(\xi(2\pi),\eta(2\pi))$  using only convergent series. We notice
that this is possible because the singularities of this mapping
appear for complex and not real values of the angles.

\ \ This method can be applied also to more complicated mappings,
like the one given by Eq. (10),because in this case as well the
singularities appear for complex and not real values of the
variables.

\ \ By using our extended method with large \emph{m} we can find very accurately the details of the asymptotic curves in the chaotic regions around the periodic orbits. By a similar method we can find the structure of the regions between the oscillating asymptotic curves (Efthymiopoulos et al. $2013$\cite{Efthym2013}) by considering cases where the integral $c=\xi^{'} \eta^{'}$ is different from zero. Thus the details of the chaotic regions around the unstable periodic orbits (and around their homoclinic points) can be obtained analytically. This shows that chaos is not at all random, but can be represented by accurate analytical formulae if we apply our new method. This finding opens new possibilities for the study of chaos in dynamical systems.

\ \ A final figure (Fig.10) shows an analytic calculation of
heteroclinic intersections. These are the intersections of the
unstable manifold of the unstable periodic orbit $(x_1=x_2=0)$ with
the stable manifold of the unstable periodic orbit $(x_1=0,x_2=\pi)$
in the standard map for $K=6$. For this value of \emph{K} the
periodic orbit $(x_1=0,x_2=\pi)$ is unstable, thus it has a stable
and an unstable manifold. The stable manifold is long enough and
intersects the unstable manifold of the orbit $(x_1=x_2=0)$ at many
heteroclinic points. In Fig.$10$ we see also homoclinic points
between the unstable manifold from the periodic orbit $(x_1=x_2=0)$
and the stable manifold from $(x_1=0,x_2=2\pi)$, which is the same
periodic orbit as $(x_1=x_2=0)$ because of the modulo $2\pi$ in the
standard map $(8)$.

\ \ The heteroclinic points are the most important manifestation of
chaos. In fact an orbit starting at an heteroclinic point of Fig.10,
tends asymptotically to the periodic orbit $(x_1=0,x_2=\pi)$ as
$t\rightarrow \infty$, while in the past it tended to the periodic
orbit $(x_1=x_2=0)$ as  $t\rightarrow - \infty$. Thus this orbit
produces a very obvious mixing of different regions of the phase
space.

\ \ The orbits shown in Fig.10 are calculated both numerically and
analytically, using the method described above, with series
truncated at order $N=20$. (The deviations between the numerical and
the theoretical curves appear after a longer time). This example of
an analytical calculation of heteroclinic orbits yields an important
new analytical result in the very heart of chaos.

\end{document}